\documentclass[11pt]{article}
\usepackage{moriond}
\usepackage{cite}
\usepackage[dvips]{graphicx}

\newcommand{\epem}   {\ensuremath{\mathrm{e^+e^-}}}
\newcommand{\znull}  {\ensuremath{\mathrm{Z^0}}}
\newcommand{\mz}     {\ensuremath{M_{\znull}}}

\newcommand{\roots}  {\ensuremath{\sqrt{s}}}
\newcommand{\oaa}    {\ensuremath{\mathcal{O}(\alpha_s^2)}}

\newcommand{\lmeff}  {\ensuremath{\Lambda_{\mathrm{eff}}}}

\newcommand{\bbbar}  {\ensuremath{\mathrm{b\overline{b}}}}
\newcommand{\dtwod}  {\ensuremath{y_{23}}}
\newcommand{\xizero} {\ensuremath{\xi^0}}
\newcommand{\qqbar}  {\ensuremath{\mathrm{q\overline{q}}}}
\newcommand{\mui}    {\ensuremath{\mu_{\mathrm{I}}}}
\newcommand{\anull}  {\ensuremath{\alpha_0}}
\newcommand{\bw}     {\ensuremath{B_W}}
\newcommand{\as}     {\ensuremath{\alpha_{\mathrm{S}}}}
\newcommand{\ddel}   {\ensuremath{\mathrm{d}}}
\newcommand{\momone}[1] {\mbox{\ensuremath{\langle#1\rangle}}}
\newcommand{\chisqd} {\ensuremath{\chi^2/\mathrm{d.o.f.}}}

\begin{document}
\vspace*{4cm}
\title{ QCD STUDIES WITH RESURRECTED JADE DATA }

\author{ S. Kluth }

\author{ for the JADE analysis group at MPI: }
\author{ M. Blumenstengel,
P.A. Movilla Fern\'andez, S. Bethke, O. Biebel, C. Pahl, J. Schieck }

\address{ Max-Planck-Institut f\"ur Physik, F\"ohringer Ring 6,
D-80805 Munich, Germany \\ skluth@mppmu.mpg.de }

\maketitle
\abstracts{
We report on recent studies of QCD performed using reanalysed \epem\
annihilation data recorded at centre of mass energies
$14\leq\roots\leq 44$~GeV by the JADE experiment operated 1979 to 1986 at the
PETRA \epem\ collider at DESY, Hamburg, Germany. The data for some
event shape observables are compared to modern Monte Carlo event
generators using parameter values obtained from LEP~1 data.  The
distribution of $\xi=\ln(1/x)$ with $x=2p/\roots$ and $p$ the momentum
of a charged hadron is measured and compared to QCD predictions.  Fits
of \oaa+NLLA (resummed) QCD predictions combined with power
corrections to event shape data including for the first time the
$\roots=14$ and 22~GeV data samples are discussed.  }

\section{ Introduction }

QCD as the theory to describe many features of hadron production in
\epem\ annihilation such as the total cross section, jet formation,
multiplicity or momentum spectra has been established mainly with data
from the experiments at the PETRA \epem\ collider, see
e.g.~\cite{naroska87,saxon88,ali88,altarelli89,maettig89}.  However,
the analyses of that time were limited by the theoretical
understanding of QCD as well as the models needed to correct for
hadronisation effects.

In particular during the LEP era from 1989 to 2000 enormous progress
in improving QCD calculations by adding higher order radiative
corrections and creating generic numerical integration
programs~\cite{yellow1qcd,event2} or resummed logarithmic
contributions (e.g.~\cite{nllathmh}) was made. Also, detailed
predictions directly relating perturbative QCD (pQCD) predictions with
hadron distributions for e.g. multiplicity or momentum spectra became
available, see e.g.~\cite{dokshitzer91,khoze97}.

The modelling of hadronisation effects had benefited from the use of
Monte Carlo event generators such as PYTHIA/JETSET~\cite{sjostrand01},
HERWIG~\cite{herwig65} or ARIADNE~\cite{ariadne3}
containing consistent descriptions of parton and hadron production
tuned to precise data from the LEP experiments, see
e.g.~\cite{yellow3qcd,yellowlep2qcdmc}.  In another approach
analytical models of hadronisation applicable to some event shape or
jet production observables had been developed allowing comparisons of
theory with data without Monte Carlo hadronisation models, see
e.g.~\cite{kluth01a}.  

PQCD effects grow with decreasing energy scale $Q$ like
$1/\ln Q$ and in particular non-perturbative effects in many
event shape or jet production observables grow with $1/Q$.  Therefore
the data from the JADE experiment can still be used for important and
unique tests of modern QCD predictions.  From another point of view
one can argue that only now our understanding of QCD and
hadronisation processes is sufficiently advanced to allow consistent
interpretation of the PETRA data.

The JADE experiment is described in~\cite{naroska87}.  The data used
in the studies presented here correspond to samples of about 1500
events at $\roots=14, 22$ and 38~GeV, about 4000 events at 44~GeV and
about 35000 events at $\roots=35$~GeV.  The JADE software consisting
of a detailed detector simulation program and the reconstruction
programs has been installed on an IBM RS6000 AIX system using the IBM
Fortran compiler~\cite{pedrophd}.  It is possible to generate final
states from \epem\ annihilation using modern generator programs like
PYTHIA/JETSET, HERWIG or ARIADNE, to pass these through the JADE
detector simulation and to finally reconstruct the output of the
detector simulation in the same way as the data.

\section{ Data vs modern MCs }

\begin{figure}[htb!]
\begin{tabular}{cc}
\includegraphics[width=0.45\textwidth]{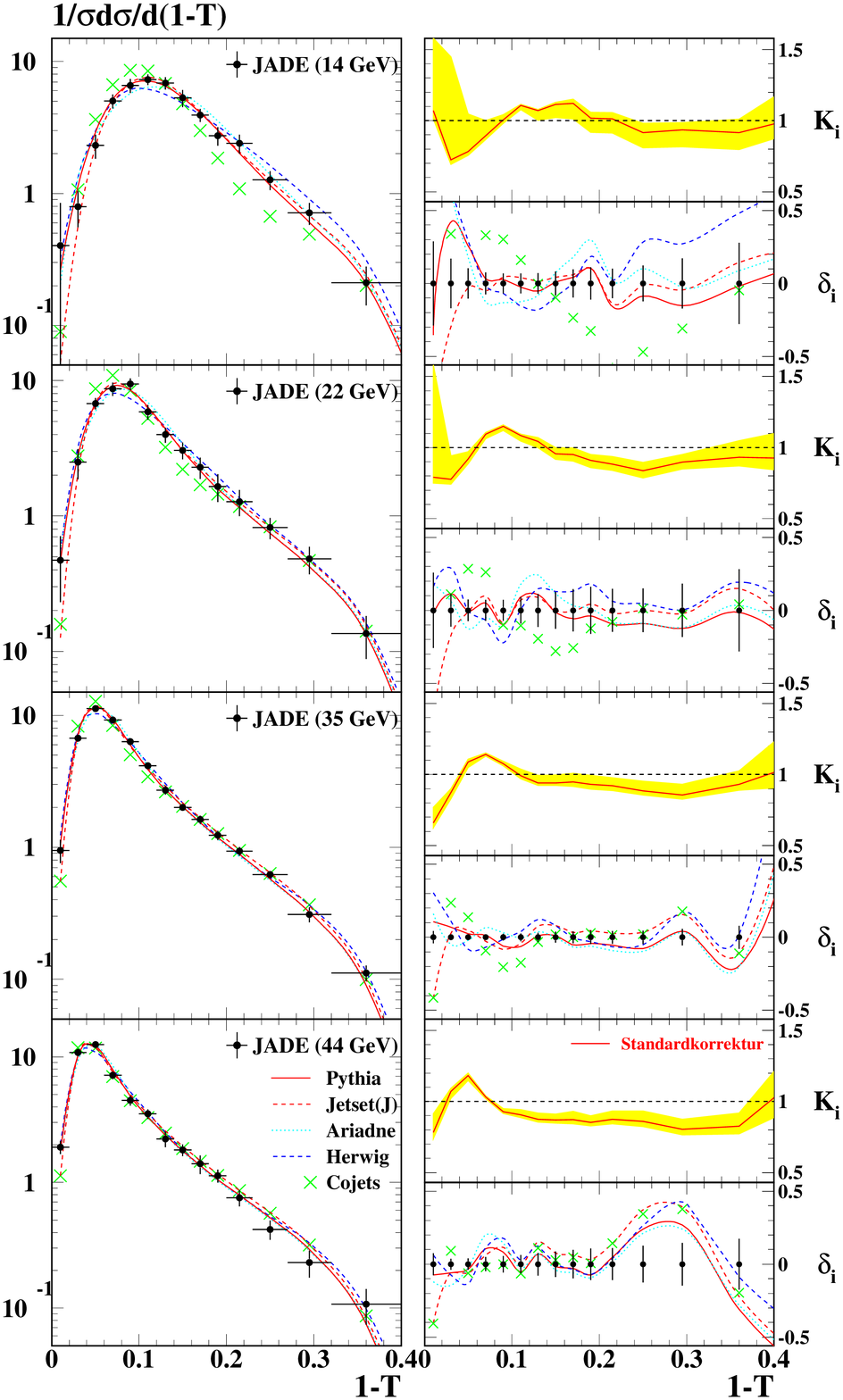} &
\includegraphics[width=0.45\textwidth]{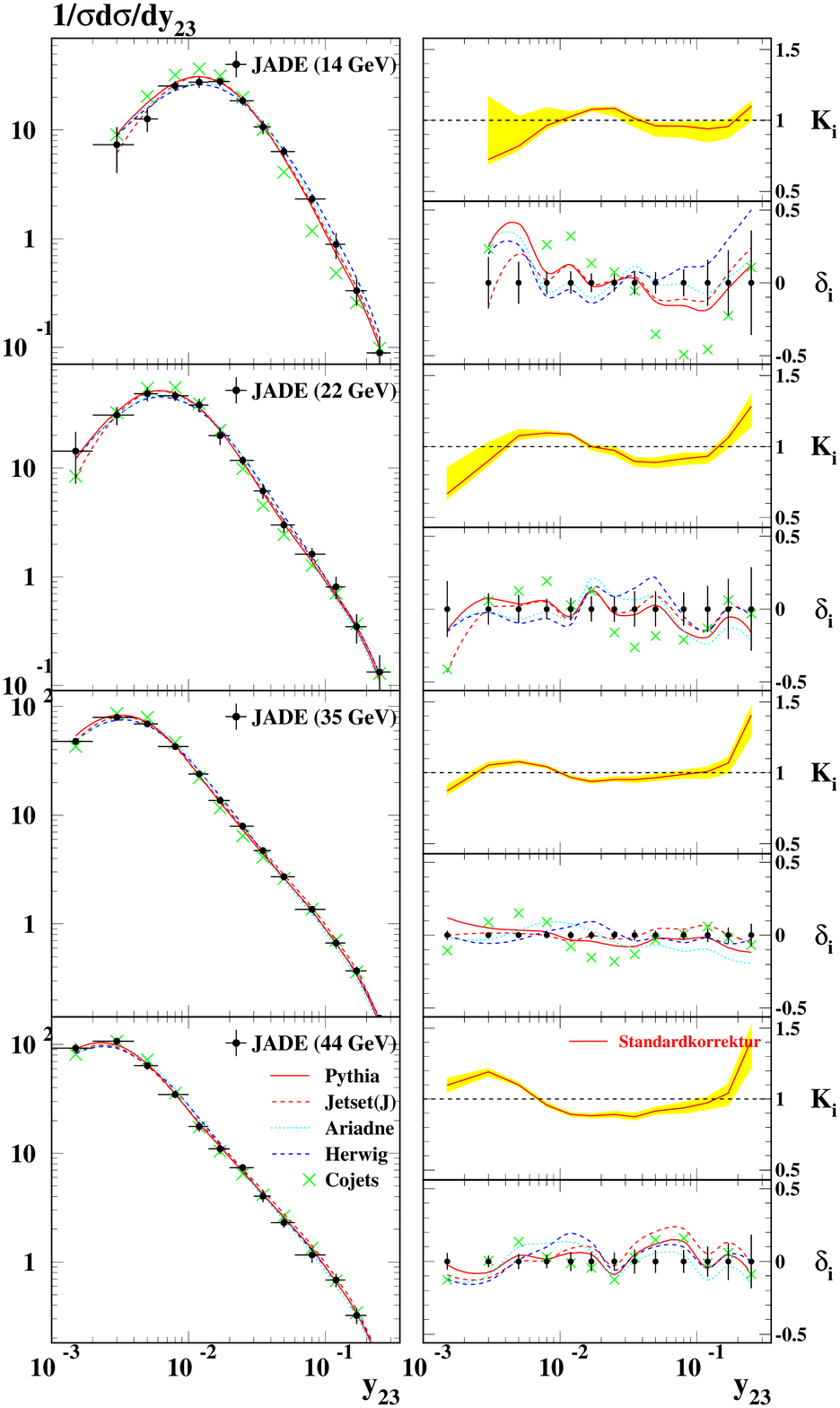} \\
\end{tabular}
\caption{ Comparison of JADE data for 1-T (left) and \dtwod\ (right)
corrected for experimental and ISR effects and \bbbar-background with
Monte Carlo predictions~\cite{pedrophd}. }
\label{fig_datamc}
\end{figure}

Before the modern Monte Carlo generators can be used with confidence
in the data analysis they must be compared with data to establish
their performance.  The data for the event shape observable thrust $T$
and the distributions of \dtwod\footnote{The value of the jet
resolution parameter where the event changes from a 2-jet to a 3-jet
configuration with the Durham jet algorithm.} shown in
figure~\ref{fig_datamc} are corrected for acceptance, detector
resolution and initial state radiation (ISR).  In addition the
contribution from $\epem\rightarrow\bbbar$ events has been subtracted
using simulated events from PYTHIA.  The $\epem\rightarrow\bbbar$
events, approximately 9\% of the JADE samples, are considered as
background since the high masses of the b-hadrons are known to distort
e.g. event shape distributions, in particular at lower
\roots~\cite{colrun,powcor,pedrophd}.  The data are compared with
predictions based on u, d, s and c quarks from the generator programs
PYTHIA~5.7, ARIADNE~4.08, HERWIG~5.9 and COJETS~6.23 as well as
JETSET~6.3 as used in earlier JADE analyses.  The parameter sets for
all programs except JETSET are taken from the OPAL collaboration,
i.e. they are derived from LEP~1 data at $\roots\simeq\mz$, while the
JETSET parameters are taken from JADE, see~\cite{pedrophd} for
details.

We find that the PYTHIA and ARIADNE generators generally describe the
JADE data well.  HERWIG is seen to produce somewhat too many 3-jet like
events and correspondingly too few 2-jet like events at lower values of
\roots\ while COJETS does not give a reasonable description of the
data.  The COJETS model implements a leading-log parton shower without
taking colour coherence into account combined with independent
fragmentation~\cite{cojets2}.  The JETSET~6.3 predictions also
describe the data fairly well.  In conclusion we find that modern
Monte Carlo Generators using parameter sets derived from LEP~1 data can be
used for QCD analyses with JADE data.

\section{ Scaled Momentum Distribution }

The shape of the distribution of $\xi=\ln(1/x)$ with $x$ being the
momentum of charged particles normalised to the beam energy is
predicted in pQCD to be approximately a skewed Gaussian, see
e.g.~\cite{dokshitzer91,khoze97}.  The position of the peak \xizero\
is predicted as a function of \roots.  The pQCD predictions in the
modified leading log approximation (MLLA) including subleading
effects~\cite{fong91,khoze97,blumenstengelphd} may be directly
compared with the hadron-level data up to a normalisation factor, a
property known as local parton hadron duality (LPHD).

\begin{figure}[htb!]
\begin{tabular}{ccc}
\includegraphics[width=0.31\textwidth]{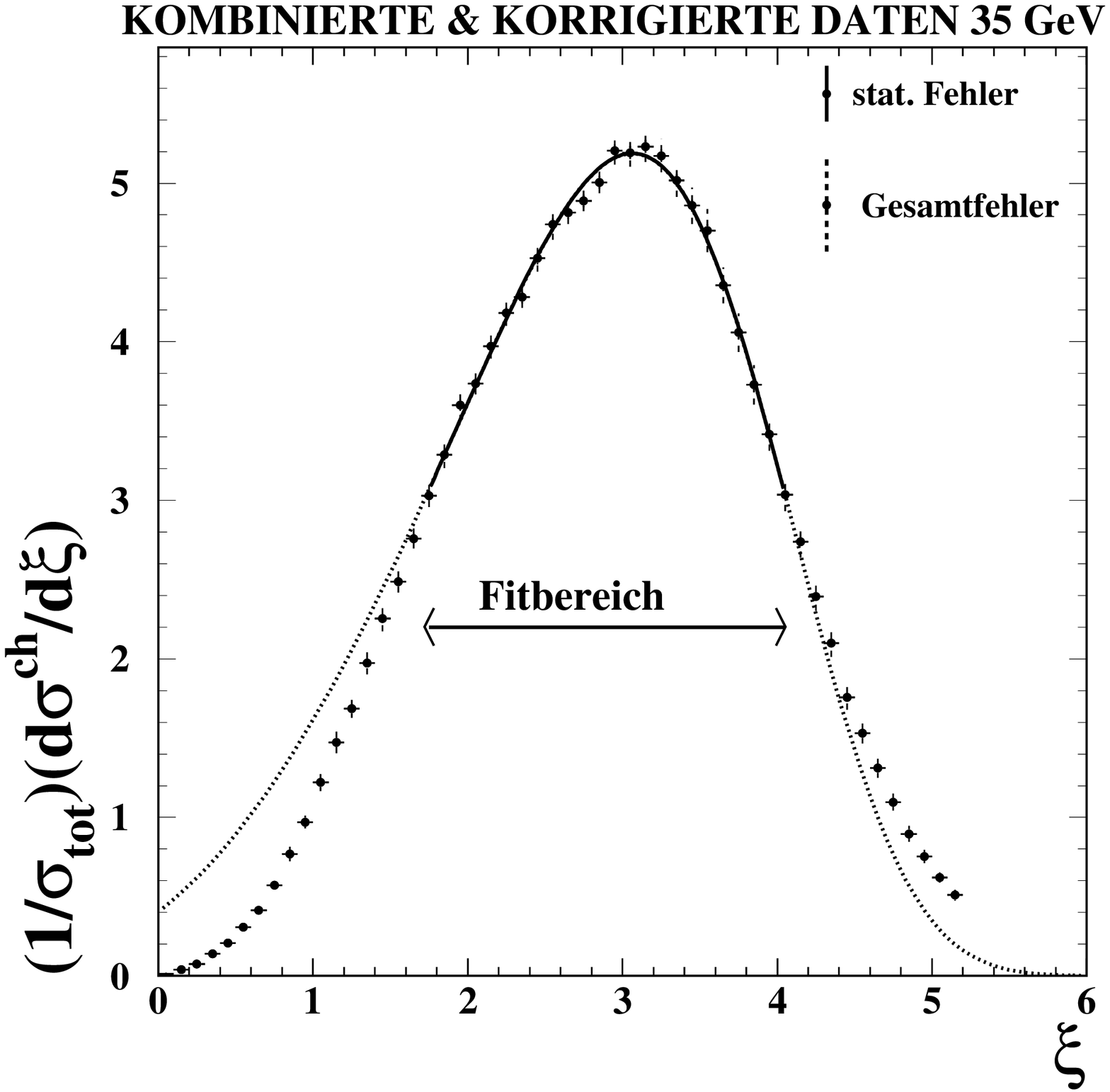} &
\includegraphics[width=0.31\textwidth]{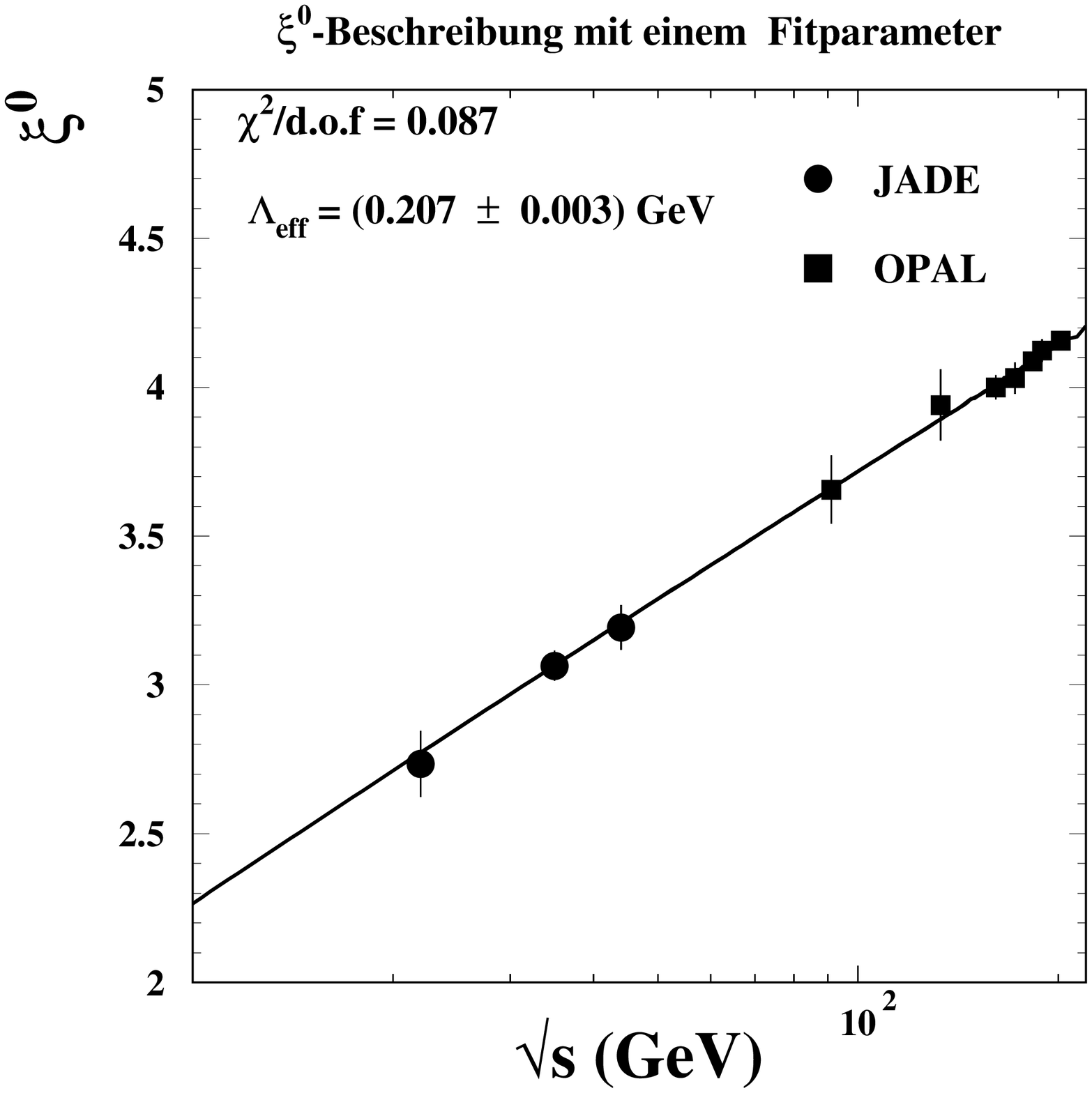} &
\includegraphics[width=0.31\textwidth]{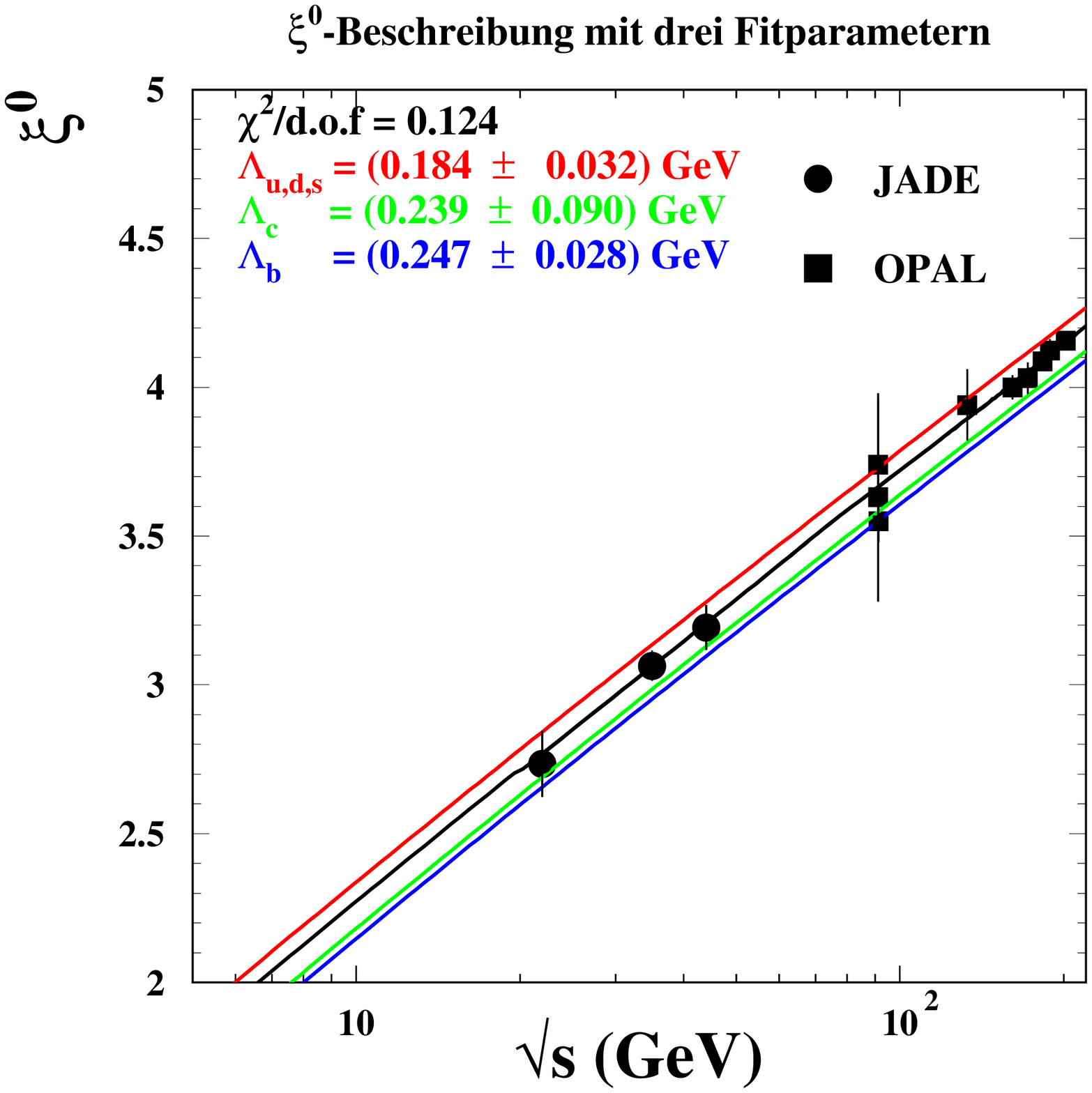} \\
\end{tabular}
\caption{ The figure on the left shows the distribution of $\xi$
measured at $\roots=35$~GeV corrected for experimental effects. The
figure in the middle shows the JADE results for \xizero\ together with
results derived from OPAL data and compared with a pQCD fit for
the \roots\ dependence of \xizero.  The figure on the right shows the
same results for \xizero\ except at $\roots\simeq\mz$ where results
from flavour separated data from OPAL are given; the fits are
explained in the text~\cite{blumenstengelphd}. }
\label{fig_mona}
\end{figure}

The $\xi$ distributions have been measured using JADE data at 22, 35
and 44~GeV~\cite{blumenstengelphd}.  In this analysis no correction
for the presence of $\epem\rightarrow\bbbar$ events has been made
consistent with analyses by other experiments.  Figure~\ref{fig_mona}
(left) shows the result for $\roots=35$~GeV.  Superimposed is a fit
of the pQCD prediction to determine the position of the peak \xizero.
The figures for the other \roots\ points at 22 and 44 GeV look similar
but have larger statistical errors due to the smaller data samples.

The fit is seen to describe the data well around the peak region; the
prediction is not expected to be in good agreement with the data
outside of the fit region.  The results for the peak positions and
effective QCD scales are given in table~\ref{tab_xi}.

\begin{table}[htb!]
\caption{ Results for peak position \xizero\ and \lmeff\ from pQCD
fits to JADE data with total uncertainties~\cite{blumenstengelphd}. }  
\label{tab_xi}
\vspace{0.4cm}
\begin{center}
\begin{tabular}{|c|c|c|}
\hline
\roots\ [GeV] & \xizero & \lmeff [MeV] \\ \hline
 22 & $2.74\pm0.09$ & $136\pm28$ \\
 35 & $3.06\pm0.05$ & $142\pm25$ \\
 44 & $3.19\pm0.06$ & $110\pm38$ \\
\hline
\end{tabular}
\end{center}
\end{table}

The dependence on \roots\ of \xizero\ is shown in
figure~\ref{fig_mona} (middle).  The three JADE data points are
compared with results derived using the same fits from OPAL data and
fitted with the pQCD prediction.  The fit describes the data
reasonably well, however, the result for \lmeff\ differs from the
results shown in table~\ref{tab_xi} by two to three standard
deviations.

In order to study effects of heavy flavours in the data samples a fit
with an effective QCD coupling \lmeff\ for each flavour (uds, c or b)
is made.  The three \lmeff\ are related by
$\xi_{\mathrm{c,b}}=\xi_{\mathrm{uds}}+0.5\ln(\Lambda_{\mathrm{c,b}}/\Lambda_{\mathrm{uds}})$
and flavour separated data from OPAL are used at $\roots\simeq\mz$. At the
other \roots\ the known flavour fractions in $\epem\rightarrow\qqbar$
are employed.  The results $\Lambda_{\mathrm{uds}}=184\pm33$~MeV,
$\Lambda_{\mathrm{c}}=239\pm90$~MeV and
$\Lambda_{\mathrm{b}}=247\pm28$~MeV indicate the presence of quark
mass effects of about 20 to 30\%; the fits are shown in
~\ref{fig_mona} (right).

\section{ Extended Power Correction Fits }

In this section we return to the study of event shape and jet
production observables, describing results of~\cite{pedrophd}.  We
concentrate on fits of pQCD predictions in \oaa\ matched with resummed
NLLA calculations as described
e.g. in~\cite{powcor,kluth01a,movilla02a}.  In these fits the
hadronisation corrections are implemented using the analytic DMW
model~\cite{dokshitzer95a}.  The strong coupling \as\ is assumed to
remain finite at very low scales below and above the Landau pole such
that it can be integrated: $\anull(\mui)=1/\mui\int_0^{\mui}\as(k)dk$
with $\mui=2$~GeV.  The value of \anull\ cannot be predicted and is a
free parameter of the model.  The main prediction of the model for
differential event shape distributions of some selected observables is
that a shifted pQCD prediction describes the corresponding
hadron-level data: $\ddel\sigma/\ddel y=(\ddel\sigma/\ddel
y)_{PT}(y-D_yP)$, where $D_y$ is an observable specific factor or
function and $P$ is an universal factor: $P\sim\mui/Q(\anull-\as)$.
Thus $P$ contains all non-perturbative dependence on the scale
$Q=\roots$ and the parameter \anull.

\begin{figure}[htb!]
\begin{tabular}{cc}
\includegraphics[width=0.5\textwidth]{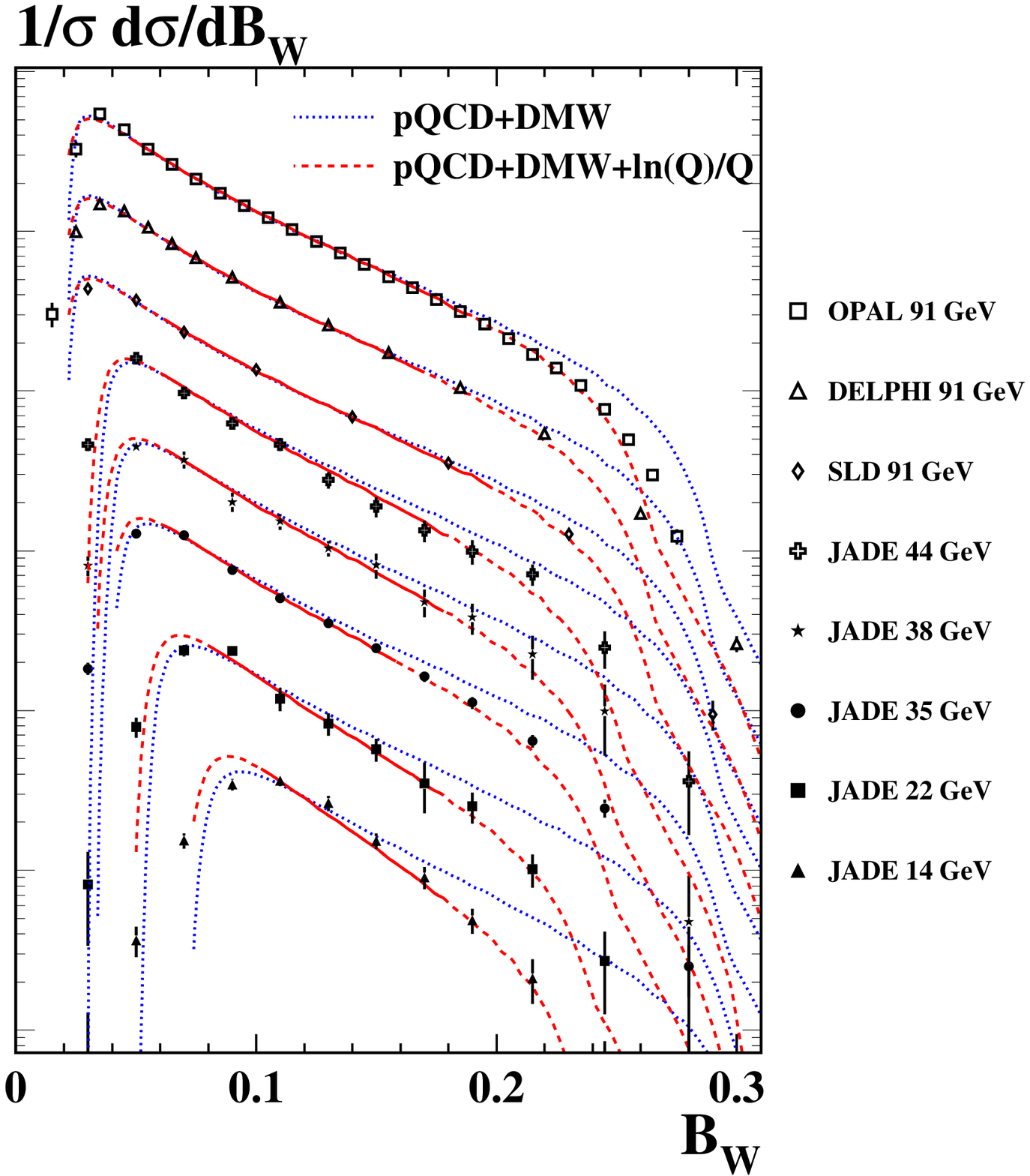} &
\includegraphics[width=0.5\textwidth]{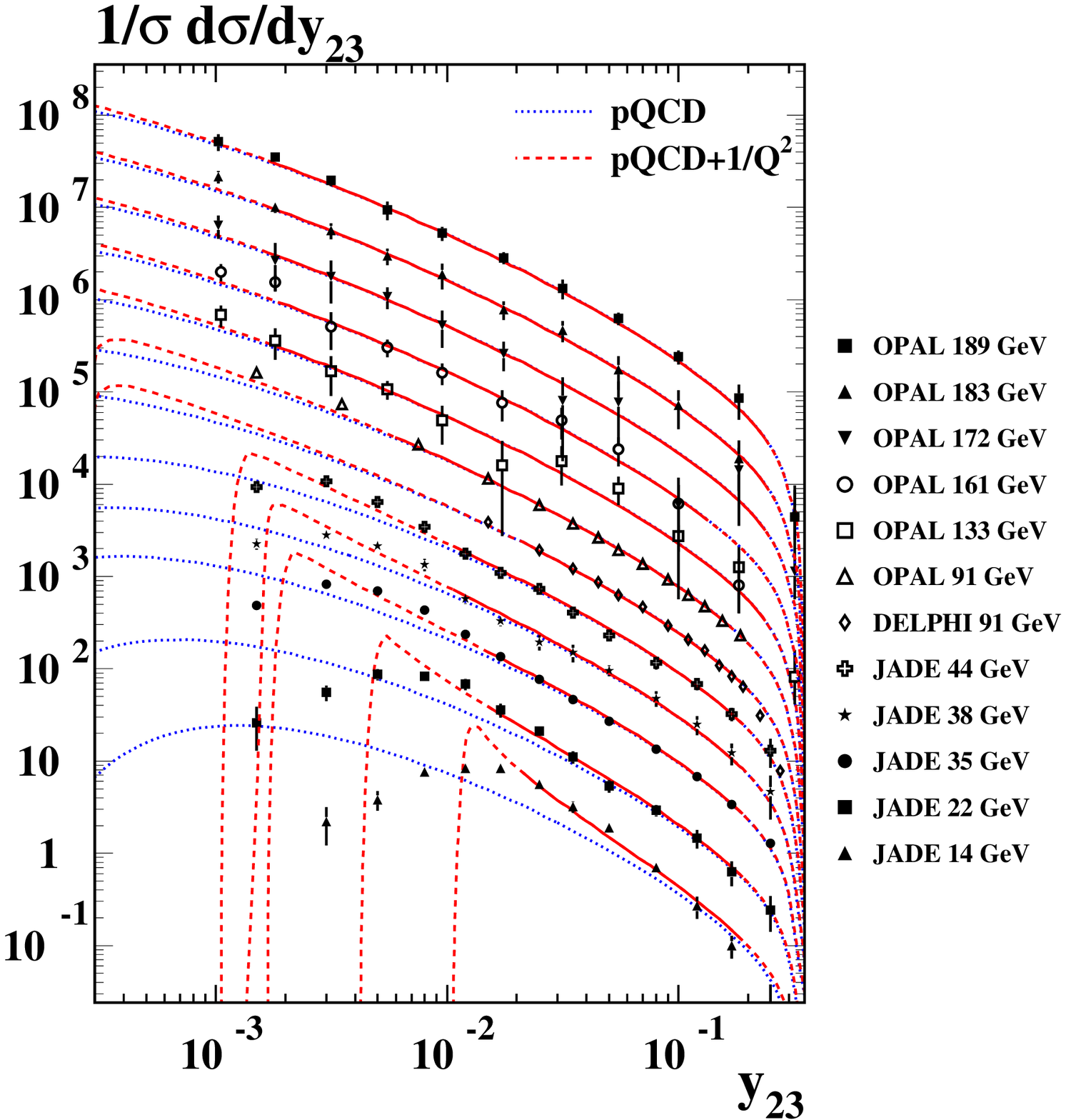} \\
\end{tabular}
\caption{ The figures show data at $\roots=91$ to 14~GeV for \bw\
(left) and \dtwod\ (right), including data from JADE.  The data are
compared with pQCD combined with standard power corrections and
modified power corrections as indicated on the
figures~\cite{pedrophd}.  The solid lines indicate the fit ranges for
all fit variations.  }
\label{fig_pedro}
\end{figure}

We want to discuss fits of such predictions to the observables \bw\
and \dtwod\ which have special properties with standard fits.  With
\bw\ global fits to data from $\roots=189$ to 14 GeV are found to
describe the data less well compared to other observables, in
particular at low \roots~\cite{pedrophd,powcor}.  Attempts were made
to add extra terms to the power correction part of the prediction and
a term of the form $A_{11}\ln(Q)/Q$ was found to provide the best
improvement, albeit with a negative coefficient
$A_{11}=-0.60\pm0.06$(fit).  The \chisqd\ value of the fit improved
from 134/154 for the standard fit to 77/153.  The results are shown in
figure~\ref{fig_pedro} (left), displaying for clarity only data at
$\roots\leq\mz$.  The dotted lines are the standard pQCD combined with
power correction fits which are seen to deviate from the data in
particular at $\roots=14$ and 22~GeV.  The fit of the same prediction
with a $A_{11}\ln(Q)/Q$ term added shown by the solid lines improves
the agreement significantly.  There is thus evidence for the presence
of additional terms in the \bw\ prediction probably behaving like
$\ln(Q)/Q$.

The DMW prediction for \momone{\dtwod}, the mean value of the
distribution, is that its power correction is proportional to
$\ln(Q)/Q^2$~\cite{dokshitzer95a}.  Previous analysis of mean values
have supported a strongly suppressed power correction~\cite{powcor}.
Here the new data for \dtwod\ from JADE are used to study the \dtwod\
distribution~\cite{movilla02b,pedrophd}.  Figure~\ref{fig_pedro}
(right) shows fits of pure pQCD to the data as dotted
lines\footnote{We don't yet use the improved numerically resummed
prediction~\cite{zanderighi}}.  The fits are seen to describe the
data well at large \roots\ but deviations become visible within the
fit ranges at low $\roots=14$ and 22~GeV.  The fit of pQCD with a
$A_{20}/Q^2$ term added is shown by the solid and dashed lines.  These
fits are indistinguishable from the pure pQCD fits at large
\roots\ but have improved agreement at low \roots.  The \chisqd\
values changed from 151/107 for the pure pQCD fit to 71/106 for the
fit with the additional $A_{20}/Q^2$ term.  The parameter is found to
be $A_{20}=2.25\pm0.18$(fit)~GeV$^2$.  The new
data at 14 and 22 GeV play an important r\^ole in discriminating
between the two possible predictions.  We thus find evidence for a
power correction probably behaving like $1/Q^2$ at low \roots\ for
\dtwod\ distributions.

\section{ Summary }

The continuing analysis of the \epem\ annihilation data of the JADE
experiment provides unique opportunities for tests of QCD.  Modern
Monte Carlo generators with parameter sets derived from LEP~1 data are
found to describe event shape and jet production observable
distributions reasonably well, with PYTHIA and ARIADNE showing a
somewhat better agreement at low \roots\ compared to HERWIG.  New
measurements of the charged particle momentum fractions $\xi=\ln(1/x)$
have been presented and compared with MLLA pQCD predictions including
subleading effects.  Studies of extended DMW power correction fits
using the new JADE data at $\roots=14$ and 22~GeV have provided
evidence for additional power correction like terms in the DMW power
correction for the observables
\bw\ and \dtwod.


\end{document}